\documentclass[prd,twocolumn,showpacs,preprintnumbers,nofootinbib]{revtex4-1}
\usepackage{graphicx}
\usepackage{dcolumn}
\usepackage{bm}
\usepackage{amsmath,amssymb,amsfonts}
\usepackage{latexsym}
\usepackage{color}

\usepackage{graphicx}
\usepackage{dcolumn}
\usepackage{bm}
\usepackage{amsmath,amssymb,amsfonts}
\usepackage{latexsym}
\usepackage{color}

\def\be{\begin{eqnarray}}
\def\en{\end{eqnarray}}
\def\non{\nonumber}

\begin{document}

\title{Scalar leptoquark effects on $B \to \mu \bar\nu$ decay}


\author{Wei-Shu Hou, Tanmoy Modak and Gwo-Guang Wong}
\affiliation{Department of Physics, National Taiwan University, Taipei 10617, Taiwan}

\begin{abstract}
Purely leptonic $B$ meson decays provide unique probes for 
physics Beyond the Standard Model. 
We study the impact of a scalar leptoquark,  $S_1$, on $B \to \mu \bar\nu$ decay.
We find that, for $m_{S_1}\sim 1$ TeV,  the $S_1$ leptoquark can 
modify the $B \to \mu \bar\nu$ rate significantly.
Such a leptoquark can in principle also alter the $B \to \tau \bar\nu$ rate. 
However, current searches from LHC and low energy physics provide 
some constraints on the parameter space.
\end{abstract}

\maketitle

\section{Introduction}
Purely leptonic $B^-$ meson decays provide clean probes to 
new physics beyond the Standard Model (SM).
The prime example is $B \to \tau\bar\nu$~\cite{Hou:1992sy}, 
where the experimental observation~\cite{Tanabashi:2018oca} has provided 
one of the strongest constraints on parameters of a charged Higgs boson,
especially in the so-called two Higgs doublet model (2HDM) 
type II~\cite{Branco:2011iw} that automatically arises with supersymmetry.
The $B\to \mu\bar\nu$ decay is further helicity suppressed,
and has not been observed so far.
It was pointed out recently that, while the ratio 
$\mathcal{B}(B\to\mu\bar\nu)/\mathcal{B}(B\to\tau\bar\nu)$ is predicted 
to be the same for both SM and the popular 2HDM type II~\cite{Hou:1992sy},  
the value could deviate~\cite{Chang:2017wpl,Hou:2019uxa} 
from the SM expectation 
in the more general 2HDM (g2HDM) that allows extra Yukawa couplings.

The Belle experiment has recently measured
${\cal B}(B\to\mu\bar\nu)
 = (6.46\pm 2.22\pm1.60)\times 10^{-7}$~\cite{Sibidanov:2017vph} 
using the full dataset of 711 fb$^{-1}$, finding $2.4\sigma$ significance. 
This should be compared with the Standard Model (SM) expectation 
around $3.92\times 10^{-7}$~\cite{Hou:2019uxa}.
In anticipation of Belle~II data, 
the measurement was further updated to 
${\cal B}(B\to\mu\bar\nu)=(5.3\pm2.0\pm0.9)\times 10^{-7}$~\cite{talk-1}, 
with the aim of improving the systematics error. 
Despite a slight drop in central value, the significance moved up 
from $2.4\sigma$ to $2.8\sigma$~\cite{talk-1}.
Assuming the SM rate, the Belle~II experiment should be able to 
observe $B\to \mu\bar\nu$ decay with $\sim 5$ ab$^{-1}$~\cite{Kou:2018nap}
in its early running. 
If the rate is actually larger than SM, observation would come sooner.

It is well known that leptoquarks (LQ) can also affect 
semileptonic and purely leptonic meson decays~\cite{Dorsner:2016wpm},
which has been of some interest lately.
The impact of a scalar LQ (SLQ) on $B\to\tau\bar\nu$ decays and the associated 
constraints have been discussed in Refs.~\cite{Dorsner:2011ai,Crivellin:2019qnh}. 
In this paper we study the effect of the SLQ, $S_1$, on $B\to\mu\bar\nu$ decay 
and discuss the possible\ constraints. 
For sake of comparison, the impact of $S_1$ on 
$B\to\tau\bar\nu$ decay is also considered. 
The experimental searches for SLQs at the LHC 
(see e.g. Refs.~\cite{Aaboud:2019bye,Sirunyan:2018btu}) are based on 
the minimal Buchm\"uller-R\"uckl-Wyler model~\cite{Buchmuller:1986zs}. 
In our study, we allow $S_1$ to couple to 
different generations of quarks and leptons~\cite{Dorsner:2016wpm}.
This work is therefore complementary to the previous study of 
Ref.~\cite{Hou:2019uxa} on $H^+$ effects in g2HDM,
where the effective 4-Fermi operator approach was adopted
to match the experimental presentation~\cite{talk-1}.
Our starting point would be from this New Physics (NP) 
Wilson coefficient language.

We find that deviations
of the $B\to\mu\bar\nu$, $\tau\bar\nu$ decay rates 
from their SM expectations are constrained in particular 
by direct searches at the LHC, as well as several low energy measurements. 
The direct search constraints arise primarily from $S_1$ pair production,
followed by $S_1 \to q \mu^\pm$ decay~\cite{Sirunyan:2018ruf,Aaboud:2019jcc}, 
which cuts into the parameter space allowed by $\mathcal{B}(B\to\mu\bar\nu)$. 
Although $\mathcal{B}(B\to\tau\bar\nu)$ is less constrained, 
the case of complex $S_1$ Yukawa couplings is constrained by 
the electric dipole moment (EDM) of the neutron~\cite{Crivellin:2019qnh}.

The paper is organized as follows. We give 
the formalism in Sec.~\ref{form}, 
then present our results on Wilson coefficients in Sec.~\ref{reslt}. 
We discuss possible constraints on SLQ Yukawa couplings in Sec.~\ref{constraints}, 
and summarize with some discussions in Sec.~\ref{disc}.

\section{Formalism}\label{form}

Let us consider the SLQ, $S_1$, which has quantum numbers
$(\bar 3,1,1/3)$ under the SM gauge group. 
The relevant Lagrangian of  $S_1$ interacting with SM quarks and leptons 
can be written as\footnote{
Concerning the stability of the proton, we turn off
the coupling of SLQ to di-quarks by imposing appropriate symmetry. 
We also do not consider right-handed neutral leptons in this paper.
}
\be
{\cal L} = \Bigl ( y^L_{ij}\,{\overline {Q^{'\;c}_{iL}}}i\tau_2 L'_{jL}
               + y^R_{ij}\,{\overline {u^{'\;c}_{iR}}}\ell'_{jR}\Bigr )S_1+H.c.
\label{eq:L1}
\en
In the above, 
$Q'_L\,(L'_L)$ denotes the left-handed quark (lepton) doublet under $SU(2)_L$, 
while {$u'_R\, (\ell'_R)$} denotes the right-handed up-type quark (charged lepton) singlet, 
and $i,\, j$ are generation indices.  
These fermion states are written in the down-type quark mass basis. 
After rotating to mass eigenbasis via the transformations 
$u'_{Li}\to(V^\dagger)_{ij}\,u_{Lj}$, $d'_{Li}\to d_{Li}$, 
$\ell'_{Li}\to \ell_{Li}$, $\nu'_{Li}\to U_{ij}\,\nu_{Lj}$, 
where $V$ and $U$ are the CKM and PMNS matrices, respectively, 
the Lagrangian of Eq.~\eqref{eq:L1} can be expanded in the mass eigenbasis 
as
\be
{\cal L}&=&
(V^*y^L)_{ij}\,{\overline {u^c_i}} L \ell_j \, S_1
-(y^LU)_{ij}\,{\overline {d^c_i }} L \nu_{j} \, S_1\non\\
&& + \; y^R_{ij}\,{\overline {u^c_i}}  R \ell_j \, S_1+H.c.,
\en
where $L,\, R = (1\mp\gamma_5)/2$.

For purely leptonic $B^-$ decays, the effective Hamiltonian  in neutrino flavor-basis is given by
\be 
{\cal H}_{\rm eff} = \frac{4G_F}{\sqrt{2}}V_{ub}
                                 \Bigl ( C_{VL}^{\ell\ell'} \mathcal{O}_{VL}^{\ell\ell'}
                                          +C_{SL}^{\ell\ell'} \mathcal{O}_{SL}^{\ell\ell'}\Bigr ),
\label{eq:Heff}
\en
where 
\be
\mathcal{O}_{VL}^{\ell\ell'} &=& \left(\bar u\gamma^\mu L b \right)
                                      \left(\bar \ell \gamma_\mu L \;\nu_{\ell'}\right),\non\\
\mathcal{O}_{SL}^{\ell\ell'} &=& \left(\bar u  L b \right)
                                     \left(\bar \ell  L \;\nu_{\ell'}\right).
\en
The SM contributes only to the $V-A$ interaction via $W$-boson exchange, 
where the Wilson coefficients are written as 
$C_{VL}^{\ell\ell'} = C_{VL}^{SM,\ell\ell'}+C_{VL}^{LQ, \ell\ell'}$, 
with $C_{VL}^{SM, \ell\ell'}=\delta_{\ell\ell'}$. 
The leptoquark contributions, at the $m_{S_1}$ scale, are given by
\be
C_{VL}^{\ell\ell'} &\simeq& \ \frac{\sqrt{2}}{8G_F V_{ub}}
                                         \frac{y^{L*}_{1 \ell} y^L_{3\ell'}}{m_{S_1}^2},\non\\
C_{SL}^{\ell\ell'} &=& -\frac{\sqrt{2}}{8G_F V_{ub}}
                                          \frac{y_{1 \ell}^{R*} y^L_{3\ell'}}{m_{S_1}^2},
\label{eq:Clk}
\en
where we have approximated the factor $V_{ud_k}y^{L*}_{k i}$, 
with $k$ summed over, as $y^{L*}_{1 i}$, 
i.e. we assume the other $y^{L*}_{2 i}$, $y^{L*}_{3 i}$ factors 
do not overpower the CKM suppression of  $V_{us}$ and $V_{ub}$, respectively.

The branching ratio for $B\to \ell\bar\nu_\ell$ in SM is well known
\be 
{\cal B}^{SM}_{B\to \ell \bar\nu_\ell}
=|V_{ub}|^2f_B^2\frac{G_F^2 m_B m_\ell^2 }{8\pi\Gamma_{B}}
\left(1-\frac{m_\ell^2}{m_B^2}\right)^2.
\en
Adding the SLQ contributions, the branching ratio 
can be expressed as 
\be 
&& {\cal B}(B\to \ell \bar\nu) =
       {\cal B}^{SM}_{B\to \ell \bar\nu_\ell}\times\non\\
&& \quad \sum_{\ell'=e,\mu,\tau} \left| \delta_{\ell\ell'}
                  - |C^{\ell\ell'}_{SL}|e^{i\phi^S_{\ell\ell'}}\frac{m_B^2}{m_{\ell} m_b}
                 + |C^{\ell\ell'}_{VL}|e^{i\phi^V_{\ell\ell'}}\right|^2,
\label{eq: B2lnu}
\en
where the undetected anti-neutrino flavor $\ell'$ in the final state is summed over.
Eq.~\eqref{eq: B2lnu} is essentially the same form used by Belle~\cite{talk-1},
except that we allow for the phase(s) $\phi_{\ell\ell'}^{S(V)}$
{of $C^{\ell\ell'}_{S(V)L}$}, which is the phase difference 
between the product of Yukawa couplings and $V_{ub}$.
Note that the Wilson coefficients corresponding to the scalar operators 
in Eq.~\eqref{eq: B2lnu} should be evolved to the $B$ meson scale 
via RGE~\cite{Tanabashi:2018oca} 
\be 
C^{\ell\ell'}_{SL}=-\frac{\overline m_b(\overline m_b)}{\overline m_b(\mu_0)}
\frac{\sqrt{2}}{8G_F V_{ub}}
\frac{y_{1\ell}^{R*}y^L_{3\ell'}}{m_{S_1}^2}\label{running},
\en
where $\overline m_b(\overline m_b)$ and $\overline m_b(\mu_0)$ 
are the $\overline{\rm MS}$ running masses
evaluated at $\overline m_b$ and $m_{S_1}$, respectively.

\begin{figure*}[htbp!]
\centering
\includegraphics[width=.44 \textwidth]{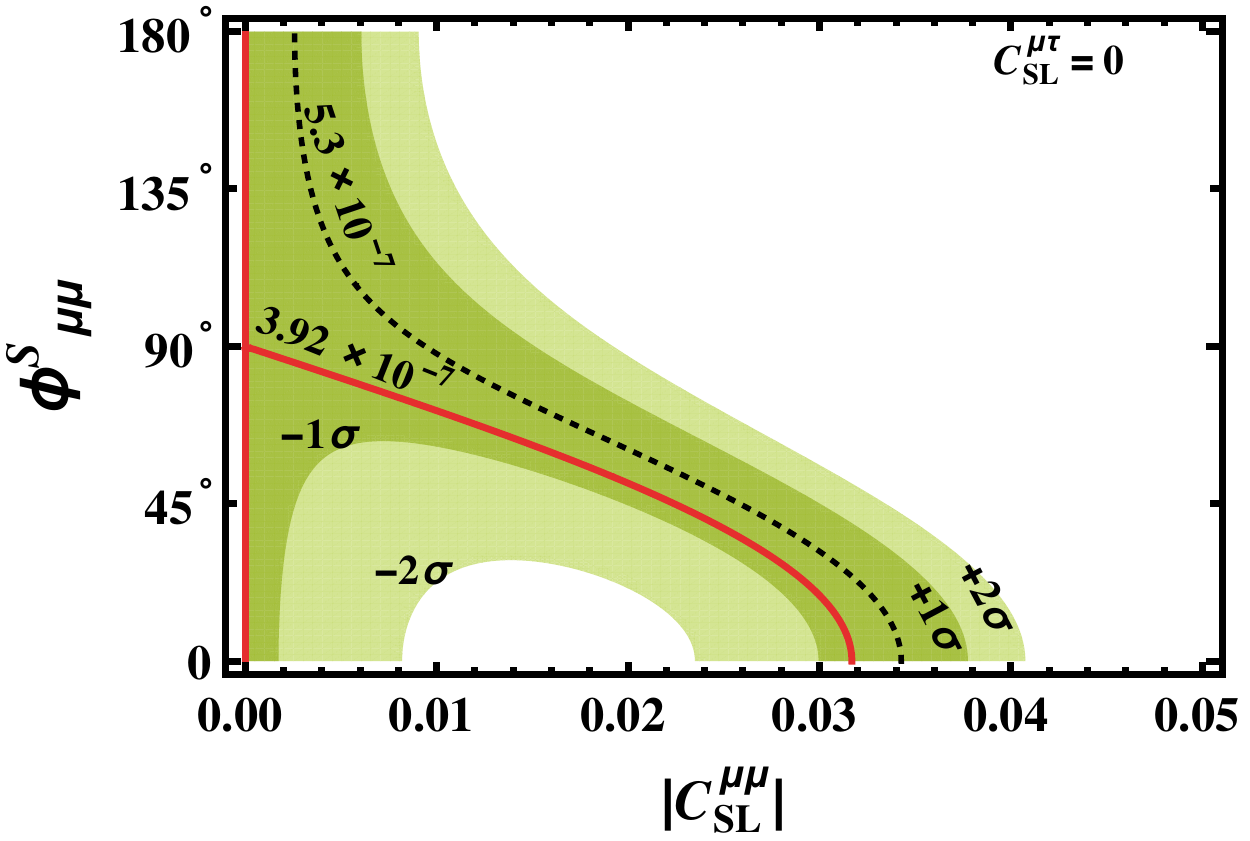}
\includegraphics[width=.43 \textwidth]{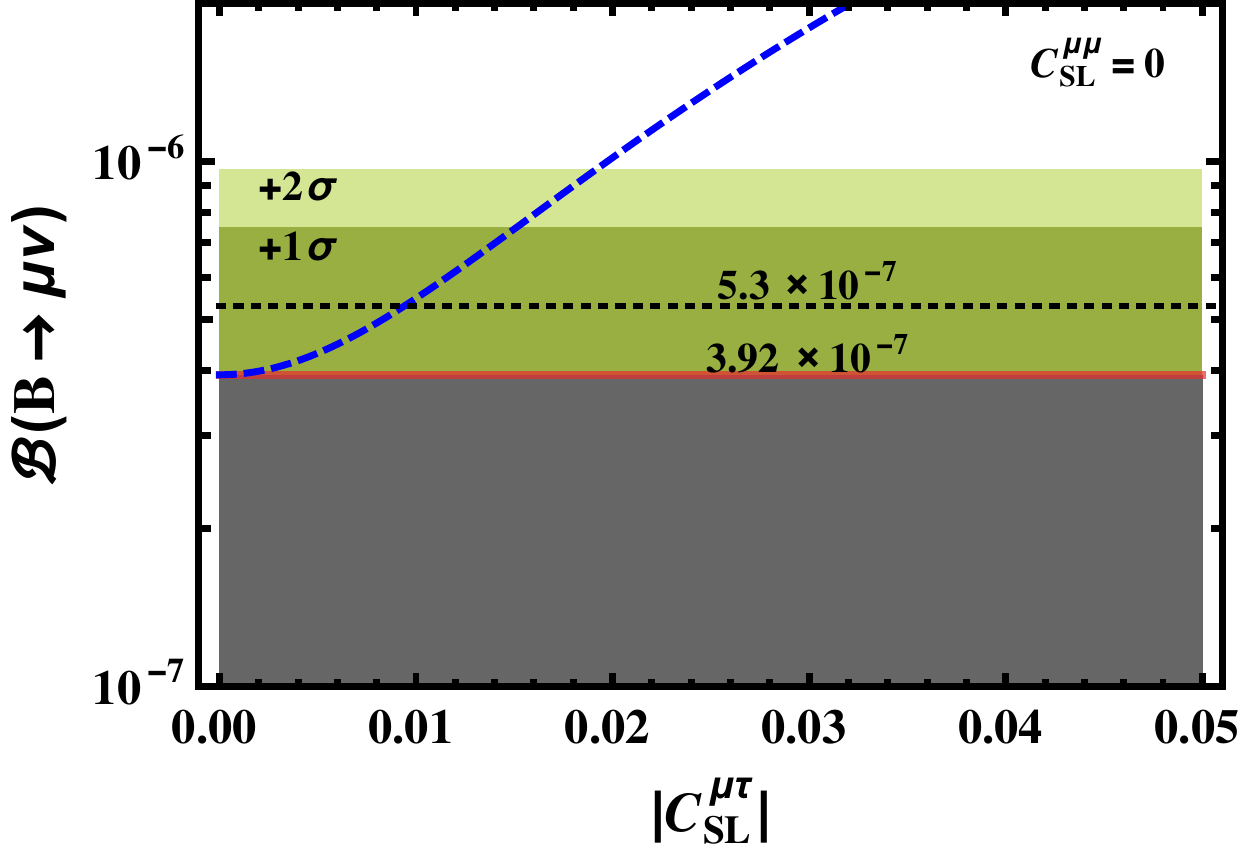}
\caption{
Branching ratio of $B\to \mu\nu$ with $|C^{\ell\ell'}_{SL}|=0$ mechanism
 ($|C^{\ell\ell'}_{VL}|=0$). 
The left panel plots the contours in $|C^{\mu\mu}_{SL}|$ vs $\phi^S_{\mu\mu}$ plane, 
and the blue dashed line in the right panel gives the 
$|C^{\mu\tau}_{SL}|$ dependence of $\mathcal{B}(B\to \mu\bar{\nu})$. 
For both panels, the red solid and black dotted lines correspond to 
the SM ($3.92\times 10^{-7}$) and Belle central values ($5.3\times 10^{-7}$), 
respectively, while the current $1\sigma$ and $2\sigma$ ranges
are shown in dark and light green shades.
The $B \to \mu\nu$ rate can only be enhanced by $|C^{\mu\tau}_{SL}|$,
where a $\bar\nu_\tau$ is emitted, which is marked by 
the darker gray shades below the SM (red) line in the right panel. 
}
\label{fig: Fig1}
\end{figure*}
\begin{figure*}[htbp!]
\centering
\includegraphics[width=.44 \textwidth]{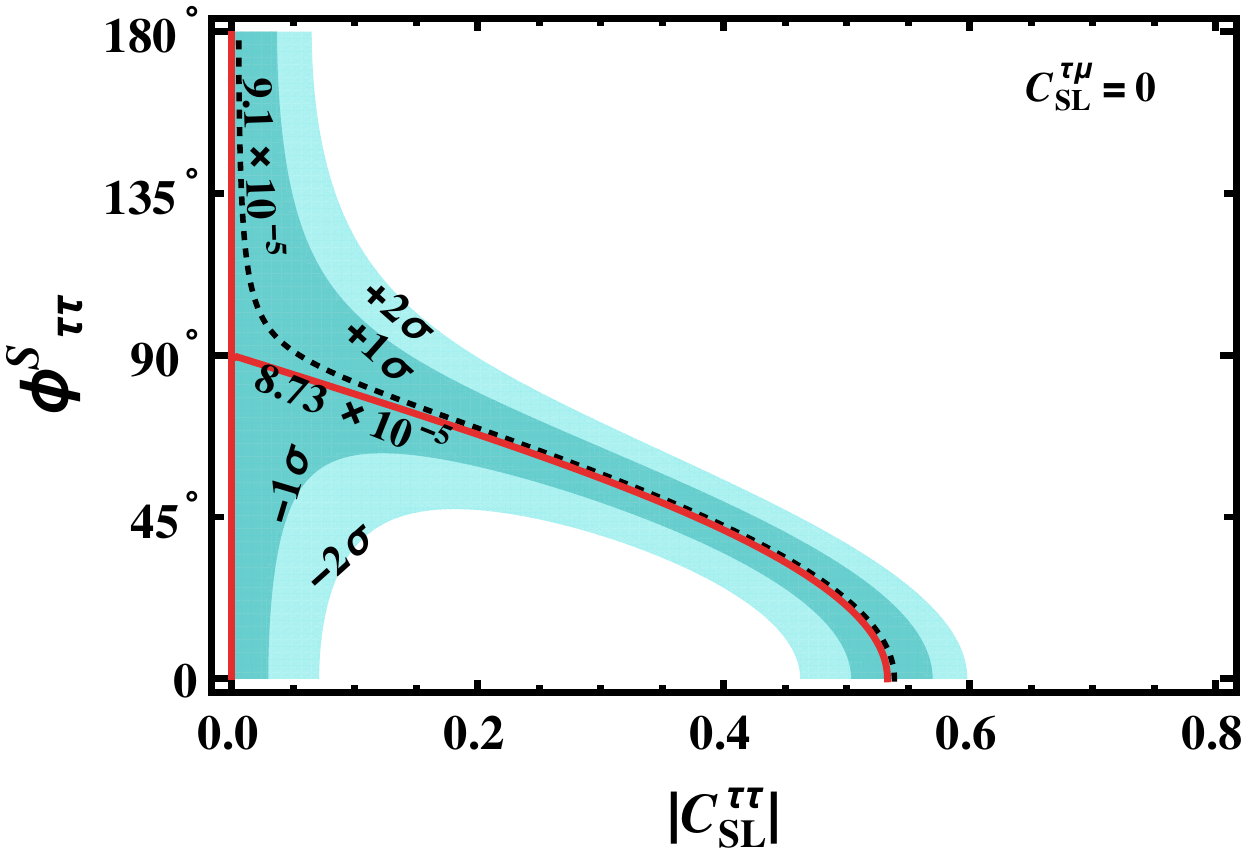}
\includegraphics[width=.43 \textwidth]{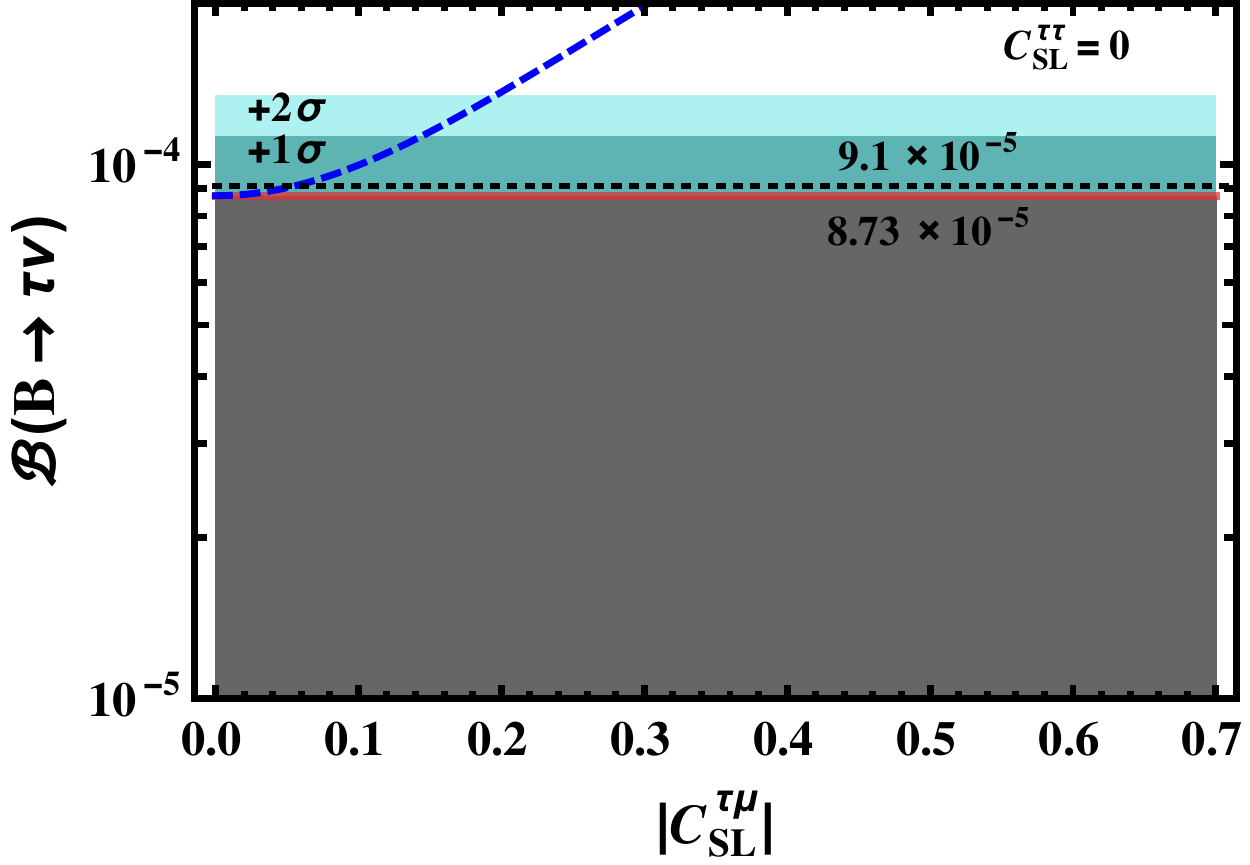}
\caption{
Similar to Fig.~\ref{fig: Fig1}, but for  $B\to \tau\bar{\nu}$ decay. 
See text for details.
}
\label{fig: Fig2}
\end{figure*}

Defining the ratio  
\be
{\cal R}_B^{\mu/\tau}\equiv
\frac{{\cal B}(B\to\mu\bar\nu)}{{\cal B}(B\to\tau\bar\nu)},
\label{eq:ratio}
\en
in SM one finds 
\be
{\cal R}_B^{\mu/\tau}|^{\rm SM}
 = \frac{m_\mu^2(m_B^2-m_\mu^2)^2}{m_\tau^2(m_B^2-m_\tau^2)^2}
 \cong 0.0045,
\label{eq:ratioSM}
\en
which is relatively precise (subject to mild QED corrections)
since it involves only lepton masses.
The prediction for 2HDM type II is the 
same~\cite{Hou:1992sy,Chang:2017wpl,Hou:2019uxa}.

The $B\to \ell\bar\nu$ ($\ell =\mu,\tau$) decay rates 
can deviate from SM predictions in the presence of the SLQ $S_1$. 
We shall ignore $y^{L(R)}_{i1}$ couplings and 
set them to zero for simplicity~\cite{Hiller:2018wbv}. 
This is similar to the treatment of Ref.~\cite{Cai:2017wry} 
where both $y^{L(R)}_{i1}$ and $y^{L(R)}_{1i}$ were assumed to be zero. 
Here we set $y^{L(R)}_{i1}=0$, 
and note that these parameters enter in the electron EDM at one loop, 
hence receive severe constraints from the recent ACME result~\cite{Andreev:2018ayy}. 
But $y^{L(R)}_{1i}$ are less constrained. 
In the following, we will first present 
the Belle measurements of $B\to\mu\bar \nu$ and $B\to\tau\bar \nu$ decays 
in terms of $C^{\ell\ell'}_{S(V)L}$ Wilson coefficients,
then discuss in some detail the constraints on relevant 
$y^{L(R)}_{1i}$ Yukawa couplings, including other possible processes.

\section{Constraining Wilson coefficients}\label{reslt}

From Eq.~\eqref{eq: B2lnu} one can see that  
both the Wilson coefficients $C^{\ell\ell'}_{SL}$ and $C^{\ell\ell'}_{VL}$
can alter $B\to \ell \bar\nu$ decay rates. 
However, the former is more efficient, 
as $C^{\ell\ell'}_{SL}$ receives the ${m_B^2}/{m_{\ell} m_b}$
enhancement factor compared with the $C^{\ell\ell'}_{VL}$ term. 
This is especially true for $B\to \mu \bar{\nu}$ decay, 
where ${m_B^2}/{m_{\mu} m_b} \sim 60$.
For $B\to \tau \bar{\nu}$ decay, 
the $C^{\tau\ell'}_{SL}$ mechanism does not 
get large enhancement because $1/m_\tau$ is smaller than $1/m_\mu$, 
but there is still some advantage over $C^{\tau\ell'}_{VL}$ 
by ${m_B^2}/{m_{\tau} m_b} \sim 4$. 
Hence, from here on we will primarily focus on $C^{\ell\ell'}_{SL}$, 
and touch only briefly on the $C^{\ell\ell'}_{VL}$ mechanism.

\subsection{$B\to \mu \bar{\nu}$ decay}

We first focus on $B\to \mu \bar{\nu}$ decay. Depending on the type of 
anti-neutrino flavor in the final state, there are different $C^{\ell\ell'}_{SL}$ Wilson 
coefficients that can modify the $B\to \mu \bar{\nu}$ rate. 
For muon anti-neutrino in final state, $C^{\mu\mu}_{SL}$ interferes  
with the SM contribution, i.e. $\delta_{\mu\mu} = 1$ in Eq.~(\ref{eq: B2lnu}), 
while $C^{\mu e}_{SL}$ and $C^{\mu\tau}_{SL}$ effects add in quadrature 
for electron and tau anti-neutrino emission. 
But as already mentioned, $C^{\mu e}_{SL}$ receives stringent constraints, 
hence we ignore this Wilson coefficient for simplicity. 

We plot $\mathcal{B}(B\to \mu \bar{\nu})$ in 
the $|C^{\mu\mu}_{SL}|$ vs $\phi^S_{\mu\mu}$ plane 
in the left panel of Fig.~\ref{fig: Fig1}, 
setting all other Wilson coefficients to zero, 
while in the right panel we give the dependence of 
$\mathcal{B}(B\to \mu \bar{\nu})$ on $|C^{\mu\tau}_{SL}|$,
where a $\bar\nu_\tau$ is emitted. 
In generating Fig.~\ref{fig: Fig1}, we used
${\cal B}(B\to\mu\bar\nu_\mu)|^{SM} \simeq 3.92\times 10^{-7}$, 
which arises from utilizing $f_B = 190$ MeV from FLAG~\cite{Aoki:2019cca}, 
and the exclusive value $|V_{ub}|^{\rm excl.} = 3.70\times 10^{-3}$~\cite{Tanabashi:2018oca}.

Taking a closer look, Fig.~\ref{fig: Fig1}(left) plots 
the contours for $\mathcal{B}(B\to \mu \bar{\nu})$ 
in the $|C^{\mu\mu}_{SL}|$ vs $\phi_{\mu\mu}^{S}$ plane.
This is because the Wilson coefficients arising from 
extra Yukawa couplings of the SLQ are in general complex. 
The SM value for $\mathcal{B}(B\to \mu \bar{\nu})$ is given by the red solid lines, 
while the central value from Belle is illustrated by the black dotted line, with 
green dark (light) shaded regions illustrating the $1\sigma$ ($2\sigma$) range~\cite{talk-1}.
For $\bar\nu_\tau$ emission, which is not distinguished by experiment,
the dependence of $\mathcal{B}(B\to \mu \bar{\nu})$ on $|C^{\mu\tau}_{SL}|$ 
is given by blue dashed line in Fig.~\ref{fig: Fig1}(right),
which can be only constructive as it adds in quadrature.
At this level of discussion, as one is using operator language
with Wilson coefficients that follow the presentation by Belle, 
the plots bear similarity to those in Ref.~\cite{Hou:2019uxa} 
that treat $H^+$ effects in g2HDM.

Before turning to $B \to \tau\bar\nu$, let us compare the
$C^{\ell\ell'}_{SL}$ and $C^{\ell\ell'}_{VL}$ mechanisms. 
We see from Fig.~\ref{fig: Fig1} that, 
to account for the Belle central value for 
$\mathcal{B}(B\to \mu \bar{\nu})$~\cite{talk-1}, one needs 
\begin{align}
 |C^{\mu\mu}_{SL}| \lesssim
0.0344\, (0.0026)\;\mbox{for}\;\phi_{\mu\mu}^S=0\, (\pi), 
\end{align}
while one finds from Eq.~\eqref{eq: B2lnu} that
\begin{align}
 |C^{\mu\mu}_{VL}| \lesssim
0.1641\, (2.1709)\;\mbox{for}\;\phi_{\mu\mu}^V=0\, (\pi),
\end{align}
which are much larger in value. Similarly, 
$|C^{\mu\tau}_{SL}| = 0.009$ is sufficient to explain the Belle central value, 
but $|C^{\mu\tau}_{VL}|$ would need to be 
0.568 to produce the same effect.
The required value of $|C^{\mu\mu}_{SL}|$ for $\phi_{\mu\mu}^S = 0\, (\pi)$ 
is about a factor of  ${m_B^2}/{m_{\mu} m_b} \sim 60$ smaller 
than that of $|C^{\mu\mu}_{VL}|$ for $\phi_{\mu\mu}^V = \pi\, (0$), 
similarly for $|C^{\mu\tau}_{SL}|$ versus $|C^{\mu\tau}_{VL}|$.
This illustrates that $C^{\ell\ell'}_{SL}$ provides a much more
efficient mechanism to modify $\mathcal{B}(B\to \mu \bar{\nu})$
compared with $C^{\ell\ell'}_{VL}$.

\begin{figure*}[htbp!]
\centering
\includegraphics[width=.44 \textwidth]{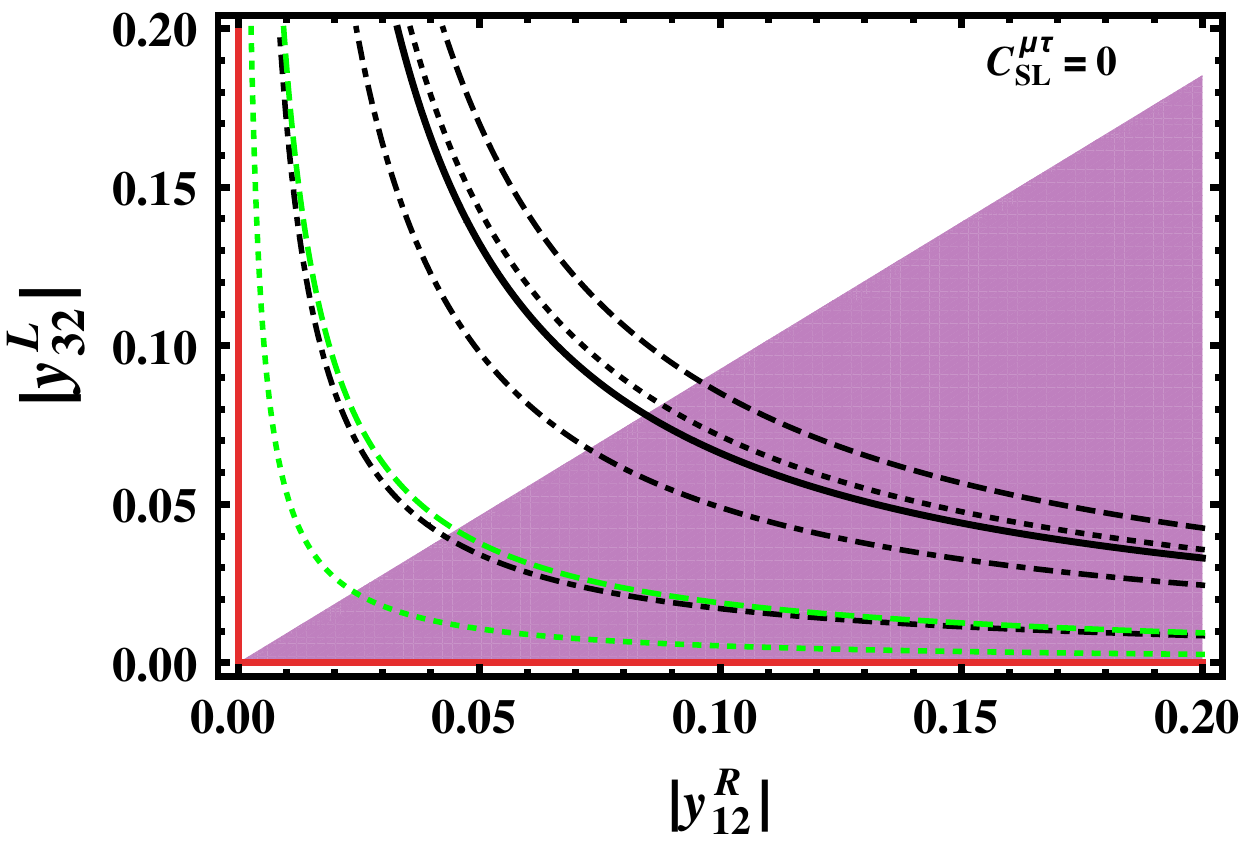}
\includegraphics[width=.44 \textwidth]{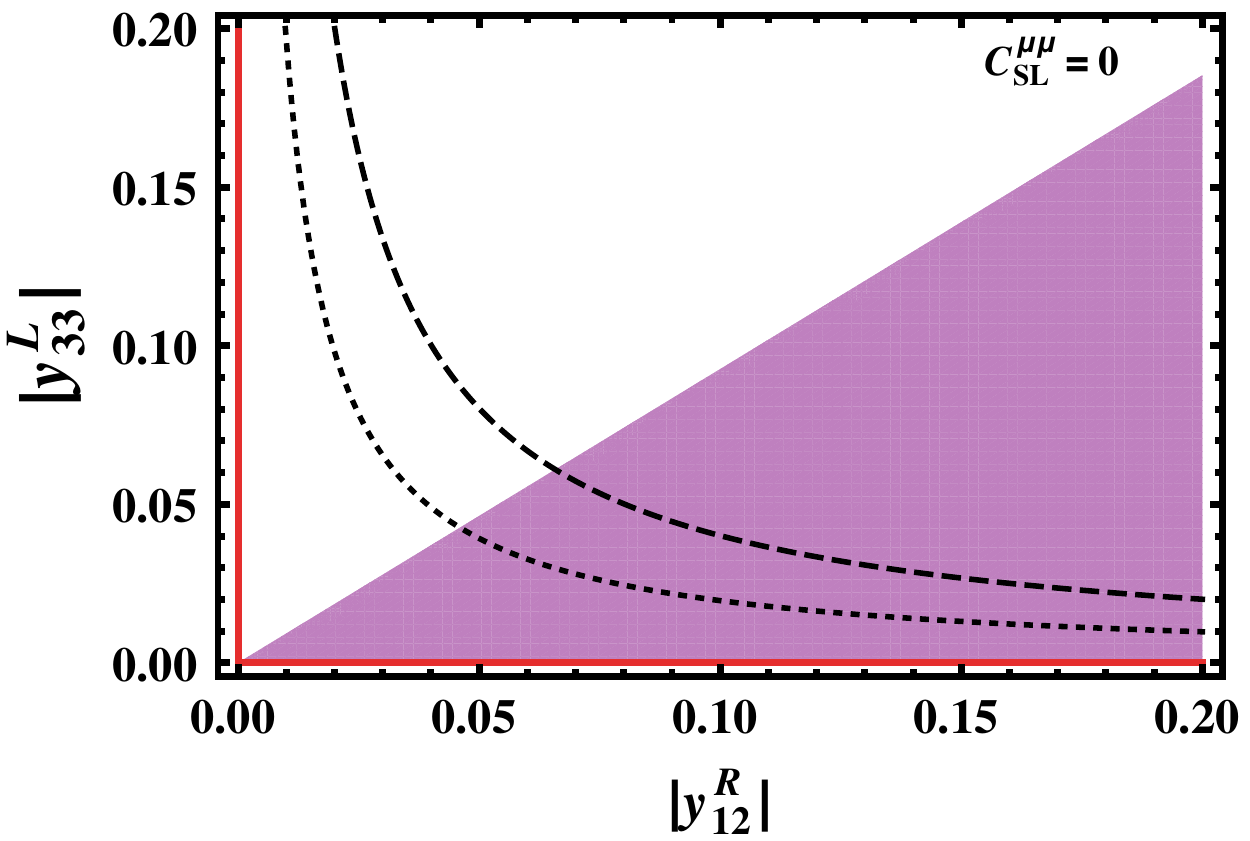}
\caption{
$\mathcal{B}(B\to \mu \bar{\nu})$ in  $|y^R_{12}|$ vs $|y^L_{32}|$ (left)
and $|y^R_{12}|$ vs $|y^L_{33}|$ (right) planes,
where we take $m_{S_1} = 1.2$ TeV. The solid contours in both panels 
represent the SM expectations, whereas the dotted, dotdashed and dashed
contours denote the central value, $-2\sigma$ and $+2\sigma$
limits of Belle measurement. In the left panel, the black and green colored contours 
are for $\phi_{\mu\mu}=0$ and $\pi$ respectively. In the right panel the 
dotted and dashed contours are plotted in black, however, it should be understood
that these contours do not depend on the overall phase, and hence black colors does not correspond to
any phase labeling.
The red solid lines in the respective panels illustrate SM expected value when the magnitude of the any 
one of the couplings is zero. The purple 
shaded regions are excluded by ATLAS SLQ pair production~\cite{Aaboud:2019jcc}.
See text for further discussion. 
}
\label{Bmuunu_yuk}
\end{figure*}
\begin{figure*}[htbp!]
\centering
\includegraphics[width=.44 \textwidth]{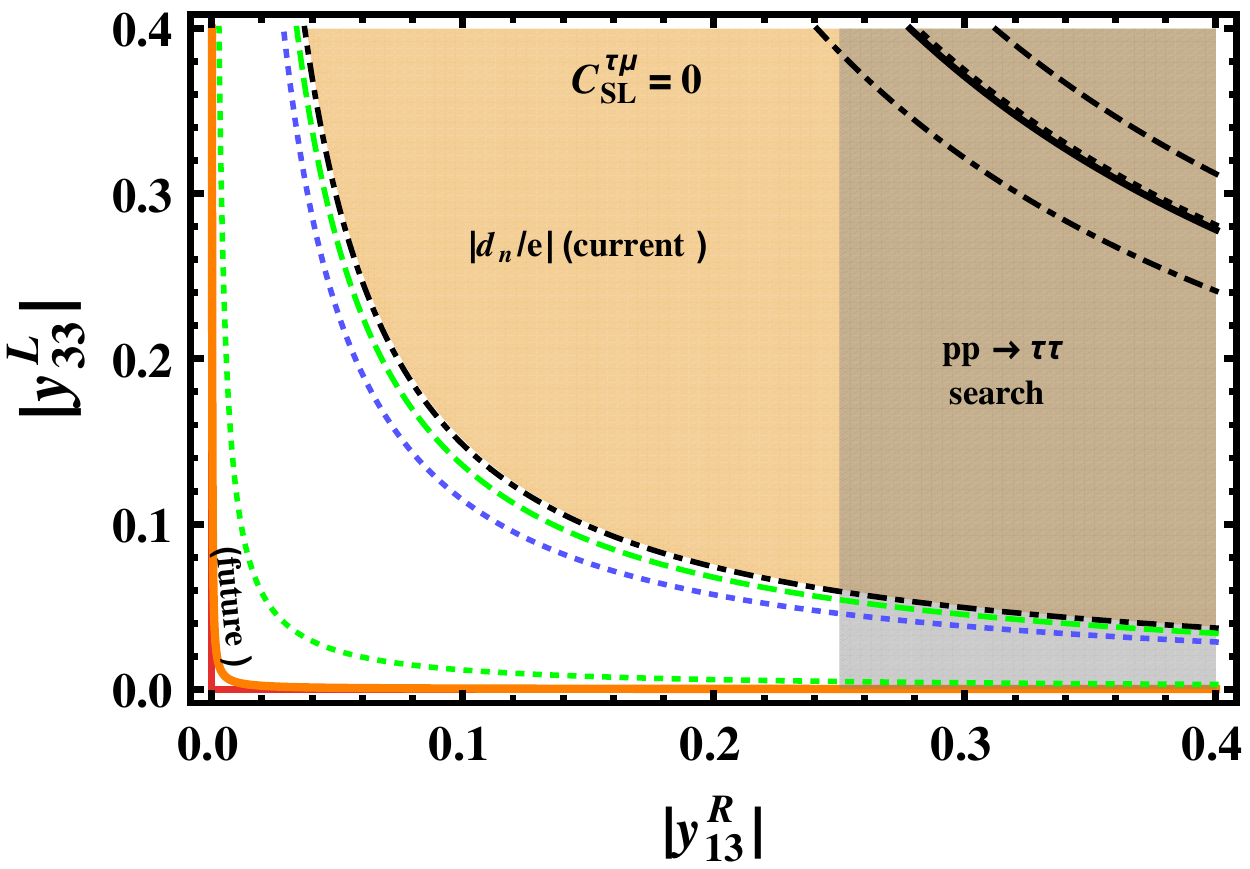}
\includegraphics[width=.44 \textwidth]{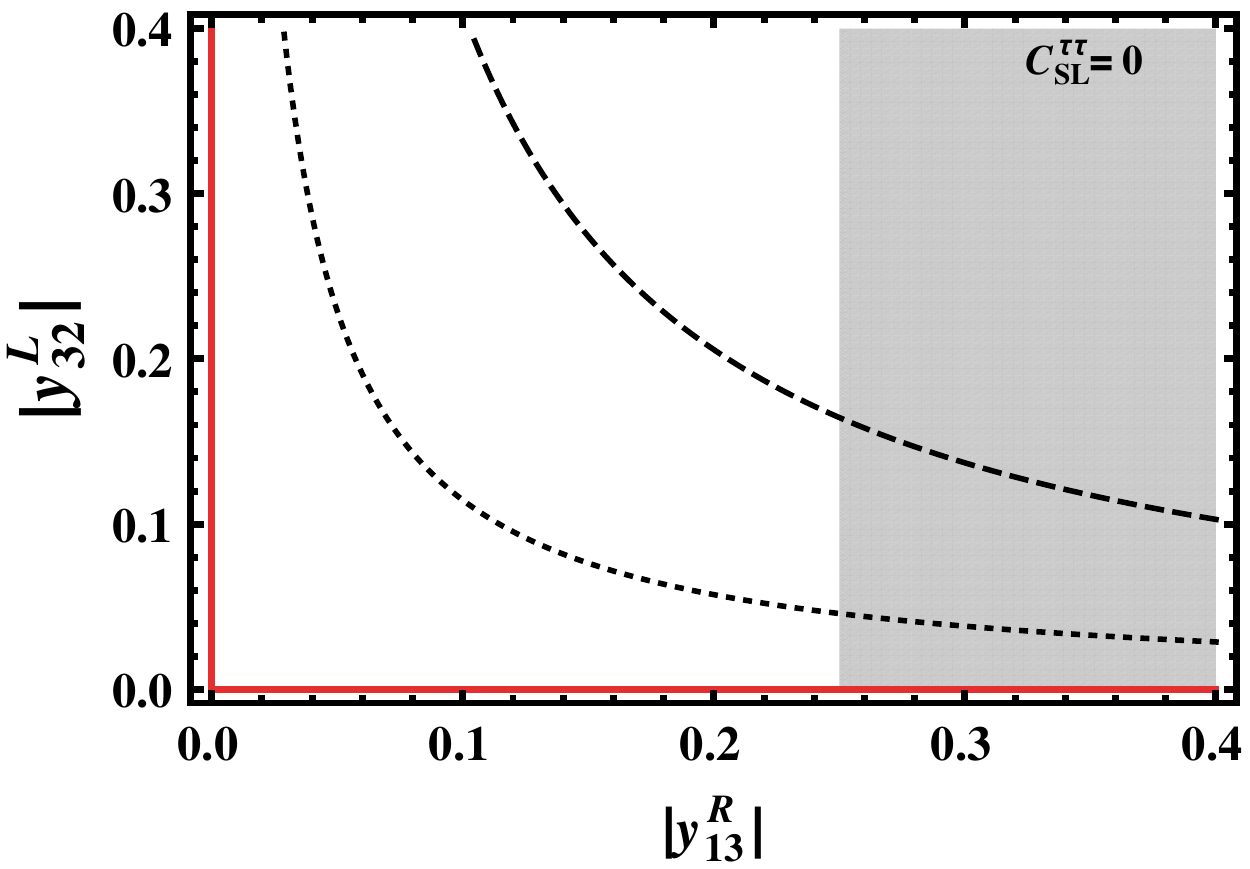}
\caption{
Similar to Fig.~\ref{Bmuunu_yuk} but for $\mathcal{B}(B\to \tau \bar{\nu})$.
The contours types and color schemes are basically same as in 
Fig.~\ref{Bmuunu_yuk} except for one additional blue dotted contour,
which represents the Belle central value for purely imaginary phase ($\phi_{\tau\tau} =\pi/2$). 
The orange shaded region and the thick orange line illustrate 
the current and future reach of constraint from neutron EDM for purely imaginary $\phi_{\tau\tau}$,
as it probes CP violation. The light gray shaded regions in both panels are 
excluded by ATLAS $pp\to \tau\tau$ search~\cite{Aaboud:2017sjh}.
See text for further discussion.
}
\label{Btauunu_yuk}
\end{figure*}

\subsection{$B\to \tau \bar{\nu}$ decay}

In Fig.~\ref{fig: Fig2}(left) we give the contours of ${\cal B}(B\to\tau\bar\nu)$  
in the $|C^{\tau\tau}_{SL}|$ vs $\phi^S_{\tau\tau}$ plane, where $\bar\nu_\tau$ is emitted, 
as well as dependence on $|C^{\tau\mu}_{SL}|$ in Fig.~\ref{fig: Fig2}(right),
for $\bar\nu_\mu$ emission.
We have used the Belle average value of 
${\cal B}(B\to\tau\bar\nu) = (9.1\pm 2.2)\times 10^{-5}$ from PDG,
while the SM expectation is $8.73\times 10^{-5}$~\cite{Hou:2019uxa}.
These are given by black dotted and red solid lines, respectively.
The $1\sigma$ and $2\sigma$ allowed ranges are illustrated by 
the dark and light cyan shaded regions, while the blue dashed line
depicts the dependence of ${\cal B}(B\to\tau\bar\nu)$ on $|C^{\tau\mu}_{SL}|$
for Fig.~\ref{fig: Fig2}(right).
Unlike $B\to \mu \bar{\nu}$ decay,
the $|C^{\tau\tau,\tau\mu}_{SL}|$ mechanisms do not have 
large enhancement factors over $|C^{\tau\tau,\tau\mu}_{VL}|$, 
as can be seen from Eq.~\eqref{eq: B2lnu}. 
As a result, larger Wilson coefficient values are needed
compared with the $B\to \mu \bar{\nu}$ case of Fig.~\ref{fig: Fig1}.

\section{Constraints on leptoquark Yukawa couplings}\label{constraints}

Our presentation so far is not so different from the $H^+$ study~\cite{Hou:2019uxa} 
in g2HDM with extra Yukawa couplings, as we follow 
the effective Hamiltonian approach used by Belle~\cite{talk-1}, 
except allowing the Wilson coefficients to be complex.
The underlying physics is, however, quite different. 
To find the allowed parameter space for $S_1$ leptoquark, 
we turn to study $\mathcal{B}(B\to \mu \bar{\nu})$ 
and $\mathcal{B}(B\to \tau \bar{\nu})$ in terms of 
the relevant $S_1$ Yukawa couplings 
$y^{R}_{12}$, $y^{R}_{13}$, $y^{L}_{32}$ and $y^{L}_{33}$.

\subsection{Constraints from $B \to \mu\bar\nu$ and $B \to \tau\bar\nu$}
For simplicity, in the left (right) panel of Fig.~\ref{Bmuunu_yuk}, 
we assume $y^R_{12}$ and $y^L_{32}$,
i.e. equivalently $y^R_{u\mu}$ and $y^L_{b\nu_\mu}$ 
($y^R_{12}$ and $y^L_{33}$, i.e. equivalently $y^L_{b\nu_\tau}$) 
are the only nonvanishing couplings.
Similarly, for Fig.~\ref{Btauunu_yuk}(left),  
we set all couplings to zero except $y^R_{13}$ and $y^L_{33}$, 
i.e. equivalently $y^R_{u\tau}$ and $y^L_{b\nu_\tau}$, whereas 
we take $y^R_{13}$ and $y^L_{32}$, i.e. equivalently $y^L_{b\nu_\mu}$ 
as the only nonzero couplings for Fig.~\ref{Btauunu_yuk}(right). 
As we focus in this subsection on the $|C^{\ell\ell'}_{SL}|$ mechanism, 
we drop the $S$ superscript from $\phi_{\ell\ell'}$ from here on.

Let us understand Figs.~\ref{Bmuunu_yuk} and \ref{Btauunu_yuk} better. 
We have chosen $m_{S_1} = 1.2$~TeV for illustration.
In both figures, the central value of Belle measurement, $+2\sigma$ and  $-2\sigma$ ranges, 
are denoted by dotted, dashed and dotdashed contours respectively. The 
solid curves illustrate the SM expectation, 
but correspond to rather sizable $S_1$ Yukawa couplings, or when one of the couplings vanishes,
which would be elucidated later.
The left (right) panel of Fig.~\ref{Bmuunu_yuk} corresponds essentially to
Fig.~\ref{fig: Fig1} left (right), where we set all $C^{\ell\ell'}_{SL} = 0$ 
except $C^{\mu\mu}_{SL}$ ($C^{\mu\tau}_{SL}$). 
However, we note that there is a negligibly small contribution (proportional to $V_{ub}$) 
in Fig.~\ref{Bmuunu_yuk}(left) from $C^{\mu\mu}_{VL}$ if $y^L_{32}$ is non-zero, 
which we neglect. A similar procedure is followed for Fig.~\ref{Btauunu_yuk}. 

For the detailed respective contours, 
in Fig.~\ref{Bmuunu_yuk}(left) we have two sets of contours for 
Belle central values of $\mathcal{B}(B\to \mu\bar\nu)$:
one for $\phi_{\mu\mu}=0$, the other for $\phi_{\mu\mu}=\pi$, 
which are denoted as black and green colors respectively.
The former requires larger values of $|y^R_{12}|$ and $|y^L_{32}|$ couplings than the latter. 
This can be understood from Eqs.~\eqref{eq: B2lnu} and \eqref{running}, where 
the $S_1$ contribution for $\phi_{\mu\mu}=\pi$ interferes constructively with SM 
in Eq.~\eqref{running}, while destructively for $\phi_{\mu\mu}=0$. 
Similarly, there are two dashed contours for Belle $+2\sigma$ range, the 
black one is associated with $\phi_{\mu\mu}=0$, while the green one is for  $\phi_{\mu\mu}=\pi$.
For Belle $-2\sigma$ range, however,
destructive interference with the SM contribution would be needed, 
which is possible only for $\phi_{\mu\mu}=0$, but not for  $\phi_{\mu\mu}=\pi$. 
In this case one has {\it two} black dotdashed contours,
illustrating the quadratic solutions (see Eq.~\eqref{eq: B2lnu}) 
for $|y^R_{12}|$ and $|y^L_{32}|$ couplings to give the Belle $-2\sigma$ range. 
A similar explanation goes for $\mathcal{B}(B\to \tau\bar\nu)$ 
in Fig.~\ref{Btauunu_yuk}(left), where there are two dotted and dashed contours 
each for the Belle central value and $+2\sigma$ range, 
corresponding to $\phi_{\tau\tau}=0\; \mbox{and}\; \pi$, shown by black and green respectively. 
But for the Belle $-2\sigma$ range there are again two black dotdashed contours 
for $\phi_{\tau\tau}=0$, due to quadratic solutions 
for $|y^R_{13}|$ and $|y^L_{33}|$ couplings.

The black and red solid contours in Figs.~\ref{Bmuunu_yuk}(left) and \ref{Btauunu_yuk}(left)  
correspond to the SM expectations for $\mathcal{B}(B\to \mu\bar\nu)$ 
and $\mathcal{B}(B\to \tau\bar\nu)$, respectively,
which require some explanation.
One would obviously recover the SM value when Yukawa couplings vanish, 
as illustrated by the red solid straight lines in each of these figures.
But for $\phi_{\mu\mu}$, $\phi_{\tau\tau}=0$ 
where the SLQ interferes destructively, 
large Yukawa couplings can overpower the SM effect
to reach the SM value, which are the black solid lines
displayed in Figs.~\ref{Bmuunu_yuk}(left) and \ref{Btauunu_yuk}(left).

In contrast, the right panels of Figs.~\ref{Bmuunu_yuk} and \ref{Btauunu_yuk} 
are straightforward: 
the SLQ contributions can only add in quadrature, 
therefore one has black dotted and dashed lines, respectively 
for Belle central values and $+2\sigma$ ranges. 

We have also plotted the Belle central value for $\phi_{\tau\tau}=\pi/2$ 
in Fig.~\ref{Btauunu_yuk}(left), to compare with the constraint from 
neutron EDM discussed later. In this case, the SLQ contribution is
purely imaginary, and adds in quadrature to the SM effect.
Note that the Wilson coefficients are generally complex,
as illustrated in Figs.~1 and 2. So, the actual interpretation
in terms of SLQ Yukawa couplings are 
more complex than what is presented here.

\subsection{$b\to c \mu \nu$ and $b\to c \tau \nu$ observables }

The presence of $y^L_{32}$ and $y^L_{33}$ can induce $b\to c \mu \nu$
and $b\to c \tau \nu$ transitions via vector Wilson coefficients
\begin{align}
 C_{VL}^{cb;\mu\mu} =& \ \frac{\sqrt{2}}{8G_F V_{cb}}
 \frac{  V_{cb} \;y^{L*}_{32} y^L_{32}}{m_{S_1}^2}, \label{bcmunu}\\
 C_{VL}^{cb;\tau\tau} =& \ \frac{\sqrt{2}}{8G_F V_{cb}}
 \frac{  V_{cb}  \;y^{L*}_{33} y^L_{33}}{m_{S_1}^2}
\end{align}
respectively.
Most notably, such Wilson coefficients contribute e.g., to $\mathcal{B}(B\to D^{(*)} \mu \nu)/\mathcal{B}(B\to D^{(*)} e \nu)$
and, $\mathcal{B}(B\to D^{(*)} \tau \nu)/\mathcal{B}(B\to D^{(*)} \ell \nu)$ (with $\ell= e,\; \mu$)
ratios respectively. The latter ratios are refereed to 
as $R{(D^{(*)})}$ ratios, where some tensions are found between the experimental measurements 
and the SM predictions~\cite{Amhis:2016xyh}.

 Let us first focus on the parameter ranges for $y^L_{33}$ in the context 
of $R{(D^{(*)})}$ anomalies.
The latest SM predictions~\cite{Amhis:2016xyh} 
are: $\mathcal{R}_{\rm {SM}}(D)=0.299\pm0.003$ and $\mathcal{R}_{\rm{SM}}(D^{*})=0.258\pm0.005$, whereas the 
world averages of the experimental measurements from HFLAV~\cite{Amhis:2016xyh} are
$\mathcal{R}(D)=0.340\pm0.027 \pm 0.013$ and $\mathcal{R}(D^{*})=0.295\pm0.011 \pm 0.008$. The 
combination of $\mathcal{R}(D)$--$\mathcal{R}(D^{*})$  world averages deviate $3.1\sigma$ from
SM predictions; which are together as the so called ``$R{(D^{(*)})}$ anomalies''. 
To find the constrain on $|y^{L}_{33}|$, we utilize
the 1D global fit value of Ref.~\cite{Blanke:2019qrx} 
which includes BaBar, Belle and, LHCb data on 
$B\to D \tau \nu$, $B\to D^{*} \tau \nu$ and  $B_c\to J/\Psi \tau \nu$ decay observables.
The Wilson coefficient $C_{VL}^{cb;\tau\tau}$ is found to be $[0.04,0.11]$ at $2\sigma$ 
for the matching scale $\mu=1$ TeV~\cite{Blanke:2019qrx}, which indicates some tension with SM.
Taking this range as ballpark value for $S_1$ LQ with $m_{S_1} = 1.2$ TeV,
we find $ 1.9\lesssim |y^L_{33}| \lesssim 3.1$ at $2\sigma$.
While finding this constraint, we assumed $b\to c e \nu$ and $b\to c \mu \nu$ transitions 
to remain SM like.

This illustrates, current 1D global fit favors rather large $|y^L_{33}|$, 
which is beyond the plotted ranges in Figs.~\ref{Bmuunu_yuk} (right) and \ref{Btauunu_yuk} (left).
However, such large $|y^L_{33}|$ would also induce $C^{\tau\tau}_{SL}$ ($C^{\mu\tau}_{SL}$) if
$y^{R}_{13}$ (or $y^{R}_{12}$) are non-vanishing and, could be sensitive to direct search limits~\cite{Greljo:2018tzh}. 
We defer discussion regarding this constraint for Sec.~4.6.
In such scenarios, to open up the parameter space for smaller $|y^L_{33}|$, 
one may require other non-zero couplings such as $y^{R}_{23}$ and $y^{L}_{23}$, or possibly 
more leptoquark couplings as discussed in Refs.~\cite{Crivellin:2019qnh,Angelescu:2018tyl}.
A detailed study with all three $y^{R}_{12}$ (or $y^{R}_{13}$), $y^{R}_{23}$, $y^{L}_{23}$ non-vanishing
couplings would be interesting in the light of new flavor and direct search results. We leave out such analysis for future.
If future measurements of LHCb and Belle-II support the   
current tension in $\mathcal{R}({D^{(*)}})$ and the anomalies become more prominent, 
the global fit value of Wilson coefficient $C_{VL}^{cb;\tau\tau}$ would deviate more from its SM prediction. 
Such large values of $C_{VL}^{cb;\tau\tau}$ could be sensitive~\cite{Greljo:2018tzh}
to direct searches at the HL-LHC (High Luminosity LHC).
On the other hand, if $\mathcal{R}({D^{(*)}})$ becomes SM like 
in future, $B\to \mu \nu$ and $B\to \tau \nu$ decays would provide sensitive probe for $|y^{L}_{33}|$
when $y^R_{12}$ and $y^R_{13}$ respectively are non-vanishing.

The ratios $R_D^{\mu/e}=\mathcal{B}(B\to D \mu \nu)/\mathcal{B}(B\to D e \nu)$ and 
$R_{D^{*}}^{e/\mu}=\mathcal{B}(B\to D^{*} e \nu)/\mathcal{B}(B\to D^{*} \mu \nu)$
are measured by Belle and found to be $0.995\pm 0.022\pm0.039$~\cite{Glattauer:2015teq} and 
$1.04\pm0.05\pm0.01$~\cite{Abdesselam:2017kjf} respectively.
Moreover, Ref.~\cite{Greljo:2015mma} found
lepton flavor universality violation between $b\to c e \nu$
and $b\to c  \mu \nu$ transitions could still be $\sim2\%$~\footnote{
Ref.~\cite{Greljo:2015mma} utilized 2014 PDG~\cite{Agashe:2014kda} fit for the 
combined $B^{\pm}/B^0$ results to estimate the lepton flavor universality violation between $b\to c e \nu$
and $b\to c  \mu \nu$ transitions. We used this value as yardstick to determine the constraint 
on $|y^L_{32}|$.}. Assuming $b\to c e \nu$ transition to be SM like,
and allowing $\sim2\%$ deviation in the $b\to c  \mu \nu$ transitions, 
we find $|y^L_{32}|\gtrsim 0.95$ is excluded for $m_{S_1}=1.2$, which is larger than the ranges plotted in Figs.~\ref{Bmuunu_yuk}.
This not surprising since the contribution from $y^L_{32}$
is suppressed compared to SM by factor $\sqrt{2}/(8G_F m_{S_1}^2)$ (see Eq.~\eqref{bcmunu}), which is about $~\sim 0.01$
for $m_{S_1}=1.2$ and induces per-mille to sub-percent level effects in $b\to c  \mu \nu$ transitions for the plotted 
range of $|y^L_{32}|$. This is also well within the $2\sigma$ allowed ranges of Belle 
$R_D^{\mu/e}$ and $R_{D^{*}}^{e/\mu}$ measurements~\cite{Glattauer:2015teq,Abdesselam:2017kjf}.

\subsection{Muon anomalous magnetic moment }

The muon anomalous magnetic moment $a_\mu$ is defined via the coupling 
$(e/4m_\mu)\,a_\mu\,\bar\mu\sigma_{\alpha\beta}\mu\, F^{\alpha\beta} $.
The $S_1$ leptoquark can generate $\Delta a_\mu$ radiatively~\cite{Cheung:2001ip,Benbrik:2010cf,Dorsner:2016wpm} via
\be
\Delta a_\mu\simeq &-&\frac{3}{8\pi^2}\frac{m_\mu^2}{ m_{S_1}^2}
\Bigl \{ \left(|V_{tb}y^{L*}_{32}|^2+|y^R_{32}|^2\right)\bigl[Q_{q^c} f_{q^c}(x_t)\bigr.\non\\
&&\bigl.+Q_S f_S(x_t)\bigr]
      - \frac{m_t}{m_\mu}{\rm Re}(V^*_{tb}y^{L}_{32}y^{R*}_{32})
        \bigl[Q_{q^c} g_{q^c}(x_t)\bigr. \non\\
&&\bigl.+Q_S g_S(x_t) \bigr]\Bigr \},
\label{eq: g-2}
\en
where $Q_{q^c}=-2/3$,  $Q_S=1/3$ for the leptoquark $S_1$,  $x_t=m_t^2/ m_{S_1}^2$, 
and the functions $f_{q^c}(x)$, $f_S(x)$, $g_{q^c}(x)$ and $g_S(x)$ 
can be found in Ref.~\cite{Cheung:2001ip}. 
This will constrain $y^L_{32}$, regardless of the value of $y^R_{32}$.

The current experimental world average~\cite{Tanabashi:2018oca} 
and the SM predicted~\cite{Keshavarzi:2018mgv} values show some deviation,
\be
\Delta a_\mu=(27.06\pm7.26)\times 10^{-10},
\en
corresponding to a long standing 3.7$\sigma$ discrepancy~\cite{Keshavarzi:2018mgv}
that could be due to New Physics.
However, for the plotted ranges in Figs.~\ref{Bmuunu_yuk} and \ref{Btauunu_yuk}, 
the contributions from $y^L_{32}$ turn out to be negligible.

\subsection{
$\tau\to\mu\gamma$ decay}

The branching ratio for $\tau\to\mu\gamma$ is given by~\cite{Benbrik:2010cf,Dorsner:2016wpm}
\be
&&{\cal B} (\tau\to\mu\gamma)\simeq
\frac{\alpha}{4\Gamma_\tau}
\frac{(m_\tau^2-m_\mu^2)^3}{m_\tau^3}
\left(|A^L_{\tau\mu}|^2+|A^R_{\tau\mu}|^2\right),\non\\
\en
where $\Gamma_\tau$ is the $\tau$ width, and 
\begin{align}
&A^L_{\tau\mu}=\frac{3}{16\pi^2 m_{S_1}^2}\times
\non\\
&\ \
\Bigl \{
(y^{R*}_{32}y^{R}_{33}m_\tau
 +|V_{tb}|^2y^{L*}_{32}y^{L}_{33}m_\mu)
   \bigl[Q_{q^c}f_{q^c}(x_t)+Q_Sf_S(x_t)\bigr]\non\\
&\quad\;
 -V^*_{tb}\;y^{L}_{33}y^{R*}_{32}\;m_t
  \bigl[Q_{q^c}g_{q^c}(x_t)+Q_Sg_S(x_t)\bigr]\Bigr \}, \\
&A^R_{\tau\mu}=\frac{3}{16\pi^2 m_{S_1}^2}\times
\non\\
&\ \  
\Bigl \{
(|V_{tb}|^2y^{L*}_{32}y^{L}_{33}m_\tau
+ y^{R*}_{32}y^{R}_{33}m_\mu)
  \bigl[Q_{q^c}f_{q^c}(x_t)+Q_Sf_S(x_t)\bigr]\non\\
&\quad\ \,
 -V_{tb}y^{L*}_{32}y^R_{33}m_t
   \bigl[Q_{q^c}g_{q^c}(x_t)+Q_Sg_S(x_t)\bigr]\Bigr \}. 
\end{align}
The current limits are
 ${\cal B}(\tau\to\mu\gamma) < 4.5\times 10^{-8}$ from Belle~\cite{Hayasaka:2007vc} 
and $4.4\times 10^{-8}$ from BABAR~\cite{Aubert:2009ag}, both at $90\%$ C.L. 
Belle II may improve the limit by a factor of 100~\cite{Kou:2018nap},
which would provide some constraint on the parameter space via $A^R_{\tau\mu}$, 
where the product $y^{L*}_{32}y^{L}_{33}$ is proportional to $m_\tau$. 
However, we find that the present constraints from Belle and Babar are again 
weaker than the range plotted in Figs.~\ref{Bmuunu_yuk} and \ref{Btauunu_yuk}.

\subsection{
EDM measurements}

The ACME experiment has put stringent~\cite{Andreev:2018ayy} constraints 
on electron EDM, $d_e$, which prompted us to set $y_{i1}^{L(R)}$ to zero.
The neutron EDM, $d_n$, imparts some constraint on the parameter space
for $\mathcal{B}(B\to \ell \bar\nu)$ decays. 

The effective Hamiltonian can be written as~\cite{Dekens:2018bci,Crivellin:2019qnh}
\begin{align}
 \mathcal{H}_{\mbox{eff}}^{\mbox{\small{EDM}}}= C_T \mathcal{O}_T+ C_\gamma \mathcal{O}_\gamma + C_g \mathcal{O}_g,
\end{align}
where the dimension-6 ${\cal O}_T$ and dimension-5 ${\cal O}_{\gamma,\, g}$
operators can be found in Ref.~\cite{Crivellin:2019qnh}.
At one loop, the leptoquark $S_1$ will contribute  to the neutron EDM
with $\tau$ and $\mu$ running inside the loop.
The contribution arising from the $\tau$ loop to the Wilson coefficients 
at the high scale can be written as
\begin{align}
 &C_T \simeq -\frac{|V_{ub}|\;|y^{L*}_{33}|\;|y^R_{13}|}{8 m_{S_1}^2} 
   {e^{-i\phi_{\tau\tau}}},\\
 &C_\gamma = -\frac{m_\tau |V_{ub}|\;|y^{L*}_{33}|\;|y^R_{13}| 
   {e^{-i\phi_{\tau\tau}}}}
 {96 \pi^2 m_{S_1}^2}\left[4+3 \log(\mu^2/m_{S_1}^2)\right],\\
 &C_g = -\frac{m_\tau|V_{ub}|\;|y^{L*}_{33}|\;|y^R_{13}|}{64 \pi^2 m_{S_1}^2} 
   {e^{-i\phi_{\tau\tau}}}, 
\end{align}
whereas the muon loop is suppressed by $m_\mu$.
Note that $V_{ub}$ enters here through the first term of Eq.~(2).

%
Neutron EDM depends on finite CPV phase.
The contribution to neutron 
EDM can be expressed as~\cite{Cirigliano:2016nyn}
\begin{align}
 &d_n/e = -(0.44\pm0.06)\,\mbox{Im}\,C_\gamma
                 -(1.10\pm0.56)\,\mbox{Im}\,C_g, \label{dn}
\end{align}
%
where $C_\gamma$ and $C_g$ are evaluated at 1 GeV, while  $C_T$ does not contribute.
We follow Ref.~\cite{Crivellin:2019qnh} for the RGE evolution
of the Wilson coefficients from the $m_{S_1}$ scale.

The current 95\% C.L. upper limit of neutron EDM,
viz. $|d_n/e| < 3.6\times10^{-26}~\mbox{cm}$~\cite{Afach:2015sja}
 (see also Ref.~\cite{Baker:2006ts}) sets strong constraint on 
the parameter space in Fig.~\ref{Btauunu_yuk}(left) for 
$\sin\phi_{\tau\tau} \neq 0$.
As illustration, we use Eq.~\eqref{dn} and find 
the orange shaded excluded region for $\phi_{\tau\tau} = \pi/2$,
i.e. purely imaginary Wilson coefficient in Fig.~2(left). 
Future measurements are expected to push the upper limit 
to $|d_n/e| < 10^{-28}~\mbox{cm}$~\cite{Hewett:2012ns}, 
which is displayed as the thick orange line. 
This illustrates that future $d_n$ measurements can exclude the whole parameter space 
of $S_1$ that supports the current Belle central value for ${\cal B}(B\to\tau\bar\nu)$, 
{\it if} $\phi_{\tau\tau} =\pi/2$, i.e. the phase of $C_{SL}^{\tau\tau}$ is near maximal.
Note that the constraint vanishes for $\phi_{\tau\tau}=0\; \mbox{or}\; \pi$, 
hence it should not be confused that the $\phi_{\tau\tau}=0\;\mbox{or}\;\pi$ contours for 
$\mathcal{B}(B\to\tau\bar{\nu})$ are excluded. 
%
The parameter space of $B\to\mu\bar\nu$ decay is less constrained 
due to $m_\mu$ suppression.

We have mainly focused on the neutron EDM. Impact  
of other EDMs such as mercury, proton, deuteron and can 
be found in more detail in Refs.~\cite{Dekens:2018bci,Crivellin:2019qnh}
for $B \to \tau\bar\nu$.

\subsection{
Direct searches}

The scalar leptoquark $S_1$ can be singly or pair produced at the LHC in $pp$ collisions 
and subsequently decay into $u_i \ell^-_j $ and $d_i {\nu}_j $ final states
 (conjugate processes are always implied), 
depending on the values of $y^R_{ij}$  and $y^L_{ij}$.
Several searches by ATLAS (e.g. Refs.~\cite{Aaboud:2019bye,Aaboud:2019jcc}) and 
CMS (e.g. Refs.~\cite{Sirunyan:2018ruf,Sirunyan:2018nkj}) 
set strong limits on leptoquark mass and branching ratios. 
At the current collision energy of $\sqrt{s}=13$ TeV, 
$S_1$ pair production via gluon fusion~\cite{Aaboud:2019jcc}
is the dominant mechanism, 
while $qg$ initiated single leptoquark production is subdominant.
For the range of $y^{R/L}_{ij}$ couplings in 
Figs.~\ref{Bmuunu_yuk} and \ref{Btauunu_yuk}, 
we find that the most relevant constraints arise from Refs.~\cite{Sirunyan:2018ruf,Aaboud:2019jcc}.

The strongest constraint comes from the ATLAS search~\cite{Aaboud:2019jcc} 
for SLQs at $\sqrt{s}=13$ TeV with 36.1 fb$^{-1}$, with final states containing 
two or more jets, one muon or electron and missing energy, 
or two or more jets with two electrons or muons. 
The search gives 95\% C.L. upper limits for branching ratios of leptoquark 
decaying into an electron and a quark, or a muon and a quark,
for different values of leptoquark masses. 
As the final state jets are not tagged, the constraint on $S_1$ parameters will
be modulated by $\mathcal{B}(S_1\to u \mu)$ if $y^R_{12}$ is nonzero.
Using the 95\% C.L. upper limit as 
$\mathcal{B}(S_1\to q \mu^\pm)$~\cite{Aaboud:2019jcc} 
for leptoquark mass of $1.2$ TeV, 
we find the purple excluded regions as displayed in Fig.~\ref{Bmuunu_yuk}.

The CMS search~\cite{Sirunyan:2018ruf} sets limit on the mass vs
branching ratio to $t\mu$ (and $t\tau$) to third generation leptoquarks.
Although only $b\nu_\ell$ type of SLQ couplings enter $B \to \ell\bar\nu$,
there are corresponding $t\ell$ couplings.
With our assumptions discussed above, for $B\to \mu \bar{\nu}$ decay of 
left (right) panel of Fig.~\ref{Bmuunu_yuk}, $S_1$ decays to 
$u \mu$, $b{\nu}_\mu$ ($b{\nu}_\tau$) and $t \mu$ ($t \tau$), 
while for $B\to \tau \bar{\nu}$ decay of left (right) panel of Fig.~\ref{Btauunu_yuk},
$S_1$ decays to $u \tau$, $b {\nu}_\tau$ ($b {\nu}_\mu$) and $t\tau$ ($t\mu$). 
With the assumed couplings, one has e.g.
$\mathcal{B}(S_1\to t \mu)\approx0.3~(0.5)$ for $y^R_{12}\simeq0.2~(0.05)$ and $y^L_{32}\simeq0.2\, (0.2)$.
For an SLQ with mass of 1.2 TeV, these branching ratios are below 
the observed~\cite{Sirunyan:2018ruf} 95\% C.L. upper limits 
at $0.56$ and $0.64$ for $t\mu$ and $t\tau$ decays, respectively.
Similarly, constraints from CMS upper limits are also 
weaker than the parameter ranges given in Fig.~\ref{Btauunu_yuk}.
The sensitivity of the HL-LHC in probing $S_1\to t \mu$ 
is discussed in Ref.~\cite{Chandak:2019iwj}, where 
$y^L_{32}\gtrsim 0.4$ is expected to be excluded at 95\% C.L.
Note that we have neglected CKM suppressed decays such as
$S_1\to c\mu$, while finding the limit on $|y^R_{12}|$.
However, the limit will weaken if $S_1$ decays to 
other final states such as $t\mu$ or $t\tau$, 
but the HL-LHC may be sensitive~\cite{Chandak:2019iwj} (see also Refs.~\cite{Mandal:2018kau,Bandyopadhyay:2018syt})
to $\mathcal{B}(S_1 \to t \ell)\sim 0.5$ (with $\ell = \mu,\, \tau$). 
The impact of direct searches with full HL-LHC dataset on the parameter space of 
$B\to\ell\bar\nu$ is worthy of further scrutiny, and will be studied elsewhere.

We remark that the couplings $y^R_{12}$ and, $y^R_{13}$ receive constraints from 
heavy resonance searches in the dilepton final states such as $pp\to\mu\mu$ (dimuon), $pp\to\tau\tau$ (ditau)
and, $pp\to\mu \tau$ via  $t$ channel $S_1$ 
exchange as discussed in Ref.~\cite{Faroughy:2016osc,Greljo:2017vvb}.
Utilizing the search for heavy resonances in the dimuon final states of Ref.~\cite{Sirunyan:2018exx}, 
Ref.~\cite{Schmaltz:2018nls} find $|y^R_{12}| \gtrsim$ 0.5 
are excluded for $m_{S_1}\sim 1$~TeV, whereas future dimuon searches 
with full HL-LHC dataset can exclude $|y^R_{12}| \gtrsim 0.1$~\cite{Raj:2016aky}.
The coupling  $y^L_{32}$ also receives constraint from such search, however, the limit 
is rather weak due to suppression from associated CKM elements.
The search for heavy resonance in the ditau final state can 
constrain $y^R_{13}$. We utilize ATLAS 
$\sqrt{s}=13$ TeV 36.1 fb$^{-1}$ ditau search result~\cite{Aaboud:2017sjh}
to constrain $|y^R_{13}|$. We closely follow the procedure outlined
in Refs.~\cite{Mandal:2018kau,Greljo:2017vvb} in our analysis for extraction of the upper limit.

The ATLAS search~\cite{Aaboud:2017sjh} is divided the into two categories, based on 
$\tau_{\rm{had}}\tau_{\rm{had}}$ and $\tau_{\rm{had}}\tau_{\rm{lep}}$ final states.
In the $\tau_{\rm{had}}\tau_{\rm{had}}$ category events with two hadronically decaying $\tau$s are 
selected, however, in the latter case events are selected such that it contains
one leptonically and one hadronically decaying tauons. The search provides
distributions for $m_T^{\rm{top}}$~\cite{Aaboud:2017sjh} (see also Refs.~\cite{Mandal:2018kau,Greljo:2017vvb} for definition)
in different bins in both the final state categories, which can be found in 
HEPData repository~\cite{hepdata}. In $pp$ collision non-zero, 
$y^R_{13}$ will induce $u\bar u \to \tau^+ \tau^-$ process via $t$-channel $S_1$ exchange and 
contribute abundantly in both $\tau_{\rm{had}}\tau_{\rm{had}}$ and $\tau_{\rm{had}}\tau_{\rm{lep}}$ categories.
As the search results in Ref.~\cite{Aaboud:2017sjh} does not veto additional activity,
we also include contributions from $ug\to S_1 \tau^+ \to \tau^+\tau^- u$ and $gg\to S_1 S_1^* \to \tau^- u \tau^+ \bar u$.
To determine the constraint on $y^R_{13}$, we generated these processes at Leading Order (LO)
in $pp$ collision at $\sqrt{s}=13$ TeV utilizing 
Monte Carlo event generator MadGraph5\_aMC@NLO~\cite{Alwall:2014hca} 
with the parton distribution function (PDF) set NN23LO1 \cite{Ball:2013hta}.
The event samples are then interfaced with 
PYTHIA 6.4~\cite{Sjostrand:2006za} for showering and hadronization, and 
finally fed into fast detector simulator Delphes~3.4.0~\cite{deFavereau:2013fsa} to incorporate detector effects.
Here we adopted the default ATLAS based detector card available in Delphes.
We adopted MLM matching scheme~\cite{Alwall:2007fs} for matrix element and parton shower merging,
and utilized the FeynRules~\cite{Alloul:2013bka} model available in Ref.~\cite{Dorsner:2018ynv}.
We have defined test statistic as~\cite{Mandal:2018kau}:
\begin{align}
\chi^2 = \sum_i \bigg(\frac{N^i_T - N^i_{\rm{obs}}}{\Delta N^i}\bigg)^2,
\end{align}
where $N^i_T = N^i_{\rm{expec.}} + N^i_{\rm{S_1}}$, $\Delta N^i = \sqrt{N^i_{\rm{obs}}}$
with $N^i_{\rm{expec.}}$, $N^i_{\rm{S_1}}$ and $N^i_{\rm{obs}}$
are the expected number of events, events from $S_1$ 
LQ and observed number of events in the $i$-th bin 
of the $m_T^{\rm{top}}$ distribution~\footnote{
In our exploratory analysis we did not add nuisance parameters
for simplicity. A more involved test statistic including
nuisance parameters can be found in Ref.~\cite{Cowan:2010js}.}.
The $\Delta \chi^2= \chi^2 - \chi^2_{\rm{min}} = 2$ corresponds
to $2\sigma$ range, with $\chi^2_{\rm{min}}$ is the 
minimum value of $\chi^2$ for $m_{S_1} =1.2$ TeV for some value 
$y^R_{13}= y^R_{13\;\rm{min}}\geq0$.
We have found $|y^R_{13}|\gtrsim 0.25$ is 
excluded at $2\sigma$. The $2\sigma$ excluded regions are shown by 
light gray shaded region in Figs.~\ref{Btauunu_yuk}.
We remark that $|y^L_{33}|$ can also be constrained 
by search in Ref.~\cite{Aaboud:2017sjh}, however the limit is 
rather weak due to suppression from CKM element $V_{cb}$.

The search for high-mass resonances in $pp \to \tau\nu$~\cite{Aaboud:2018vgh,Sirunyan:2018lbg} can 
constrain $|y^L_{33}|$ due to the presence of $c\tau S_1$ and $b\nu S_1$ couplings, as discussed in Ref.~\cite{Greljo:2018tzh}.
The limit is not stringent due to weak $c$ and $b$ PDF, however, the excluded
regions lie beyond the plotted ranges in Fig.~\ref{Bmuunu_yuk} (right) and Fig.~\ref{Btauunu_yuk} (left).
Such a search can also constrain coupling products $|y^{R*}_{13}\;y^L_{33}|$.
Recasting the $2\sigma$ upper bound of the Wilson coefficient $|C^{\tau\tau}_{SL}|$ (at $m_b$ scale) 
of Ref.~\cite{Greljo:2018tzh}, we found $|y^{R*}_{13}\;y^L_{33}| \gtrsim 0.28$ is excluded, whereas, the full Run-2 
dataset can exclude $|y^{R*}_{13}\;y^L_{33}| \gtrsim 0.17$. 
As the neutrino flavor is not measured similar upper bound can also be 
set on $|y^{R*}_{13}\;y^L_{32}|$. 
In addition, search for heavy resonance decaying into $\mu\nu$ final state (denoted as $pp\to\mu\nu$ search)~\cite{Aad:2019wvl}
can constrain $|y^R_{32}|$ as well as coupling products $|y^{R*}_{12}\;y^L_{32}|$ and $|y^{R*}_{12}\;y^L_{33}|$. 
However, we find these constraints to be weaker and excluded regions lie just outside the plotted
ranges of Fig.~\ref{Bmuunu_yuk}. As discussed earlier, current 1D global fit of $b\to c \tau \nu$ observables
favors rather large $|y^L_{33}|$,
which means parameter space of $B\to \tau \bar \nu$ in Fig.~\ref{Btauunu_yuk} (left) 
($B\to \mu \bar \nu$ in Fig.~\ref{Bmuunu_yuk} (right))
where $|y^{R*}_{13}\;y^L_{33}|$ ($|y^{R*}_{12}\;y^L_{33}|$) product is somewhat large could be
excluded by $pp\to \tau \nu$ ($pp\to \mu \nu$) search with full Run-2 or early Run-3 data. 
Furthermore, search for heavy resonances decaying into 
the $\tau\mu$ final state~\cite{Aaboud:2018jff} can potentially 
constrain coupling products such as $|y^{R*}_{12}\;y^L_{33}|$ 
and $|y^{R*}_{13}\;y^L_{32}|$ (see Eq.~\eqref{eq:L1}), however,
we find the limits to be not yet relevant for the coupling ranges 
considered in this paper (see also Ref.~\cite{Bansal:2018eha} for similar discussion).

\section{Discussion and Summary}\label{disc}

We offer a few brief remarks in passing.
The decays of $Z$ and $W$ bosons can constrain the parameter space for $y^L_{ij}$.
For example, $Z \to \tau \tau$ and $W \to \tau \nu$ exclude 
$y^L_{33}\gtrsim 1$ at $2\sigma$ for $m_{S_1}\approx 1$ TeV~\cite{Crivellin:2019qnh}.
Such constraints are, however, in general weaker than the ranges 
plotted in Figs.~\ref{Bmuunu_yuk} and \ref{Btauunu_yuk}. 
Note that the effect of $S_1$ on $B \to\tau\bar\nu$ has been 
previously discussed~\cite{Crivellin:2019qnh},
and our discussion is only for comparison with $B\to \mu\bar\nu$.
In principle, other SLQs such as $S_{3}$, $R_{2}$, $\tilde{R}_{2}$
(see Ref.~\cite{Dorsner:2016wpm} for definition), 
as well as vector leptoquarks $U_1$, $U_3$, $V_2$, $\tilde{V}_{2}$ 
can all potentially affect $B\to\ell\bar\nu$ decays. 
We leave a detailed study of these for the future.

In some sense, it is remarkable that somewhat large leptoquark Yukawa couplings
as displayed in Figs.~\ref{Bmuunu_yuk} and \ref{Btauunu_yuk} remain unexplored.
We have seen that, when light jets are involved, 
ATLAS data~\cite{Aaboud:2019jcc} provide strong constraints.
Moreover, direct searches such as $pp\to \mu\mu$ and $pp\to \tau\tau$ provides
meaningful constraints on the available parameter space for both
$B \to \mu\bar\nu$ or $B \to \tau\bar\nu$ decays.
But this also illustrates the relative arbitrariness of the $S_1$ scalar leptoquark,
where putting the mass above TeV scale on one hand escapes LHC detection,
but on the other hand demands the rather large Yukawa couplings 
(the bottom Yukawa coupling is $\sim 0.02$ in SM)
to have an effect on purely leptonic $B^-$ decays.
In contrast, the $H^+$ effect of the general 2HDM that allows for 
extra Yukawa couplings is much more nuanced~\cite{Hou:2019uxa}.
The charged Higgs could be sub-TeV, with rather weak extra Yukawa couplings,
but could still enhance $B\to \mu\bar\nu$ (less so for $B \to \tau\bar\nu$)
within the Belle allowed range.

In summary, we have explored the constraints placed by current Belle results 
on the Wilson coefficients that can affect $B\to \mu\bar\nu$, $\tau\bar\nu$ decays,
and then interpreted in terms of the Yukawa couplings of the $S_1$ scalar leptoquark.
With $m_{S_1}$ set at 1.2 TeV, rather sizable Yukawa couplings are needed
for enhancing the purely leptonic $B^-$ decays to the 2$\sigma$ upper reach
of Belle measurements.
As one awaits eventual Belle II observation of $B \to \mu\bar\nu$
and improved measurements of $B\to \tau\bar\nu$, we find that
neutron EDM can probe the CP violating phases of 
$S_1$ Yukawa coupling, while a large part of the rather large 
leptoquark Yukawa coupling range remains to be explored at hadron colliders.

\vskip0.5cm
\noindent{\bf Acknowledgments.} \
TM thanks Tanumoy Mandal for discussion.
This work is supported by grants MOST 106-2112-M-002-015-MY3, 
108-2811-M-002-537, 107-2811-M-002-039, and NTU 108L104019.

\end{document}